\documentclass[epj
]{svjour}
\usepackage{graphics}
\usepackage{amsmath}
\usepackage{rotating}
\begin{document}
\title{Spectator-model operators in point-form relativistic quantum
mechanics}
\author{ T.~Melde\inst{1} \and L.~Canton\inst{2} \and W.~Plessas\inst{1}
\and R.F.~Wagenbrunn\inst{1}
}                     
%
%
\institute{
Theoretische Physik, Institut f\"ur Physik, Universit\"at Graz,
Universit\"atsplatz 5, A-8010 Graz, Austria
\and 
INFN, Sezione di Padova, and Dipartimento di Fisica, 
Via Marzolo 8, I-35131 Padova, Italy
}
\date{Received: date / Revised version: date}
%
\abstract{
We address the construction of transition operators for 
electromagnetic, weak, and hadronic reactions of relativistic
few-quark systems along the spectator model. While the problem is of
relevance for all forms of relativistic quantum mechanics, we
specifically adhere to the point form,  
since it preserves the spectator character of the corresponding
transition operators in any reference frame. The conditions imposed
on the construction of point-form spectator-model operators are
discussed and their implications are exemplified for mesonic decays of
baryon resonances within a relativistic constituent quark model.
} 
\titlerunning{Spectator model operators in point-form relativistic quantum
mechanics}
\maketitle

\section{Introduction}

Relativistic quantum mechanics (RQM) has been known for a long time
as a feasible method to treat multi-particle systems
in a Poincar\'e-invariant way~\cite{Keister:1991sb}. Already in 1949
Dirac described the front, instant, and point forms of relativistic
dynamics~\cite{Dirac:1949}. Based on a complete classification of
subgroups of the Poincar\'e group, Leutwyler and
Stern~\cite{Leutwyler:1977vy} also considered (the only) two further
classes of RQM with a maximal transitive stability group. Most of the
practical investigations in hadronic physics have been performed in
the front and instant forms. Only in recent years the point form has
attracted increasing attention. For instance, it has been applied
to calculate the $\pi$ charge form
factor~\cite{Allen:1998hb}, electromagnetic baryon form
factors~\cite{Coester:1997ih,Berger:2005}, electroweak nucleon form
factors~\cite{Wagenbrunn:2000es,Glozman:2001zc,Boffi:2001zb,Berger:2004yi},
and also widths of $\pi$ and $\eta$ decay modes of $N$ and $\Delta$ 
resonances~\cite{Melde:2002ga,Melde:2004xj}. In all cases a 
point-form spectator model (PFSM)~\cite{Klink:1998pr}
has been adopted for the current and decay operators, respectively.

Tackling the full theory of a relativistic three-body system,
including all genuine many-body operators required by Poincar\'e
invariance (as well as hermiticity and current conservation) has not
yet been possible in any one of the different forms of RQM
mentioned above.
Therefore one has resorted to simplified transition operators congruent
with an impulse approximation in a nonrelativistic theory. However,
this practice causes problems in RQM, since the constraints that have
to be imposed on covariant relativistic operators lead to serious
difficulties in the definition and application of one-body
operators~\cite{Polyzou:1986df,Sengbusch:2004sf}.
Consequently, the constructions of so-called spectator-model
operators are afflicted with specific shortcomings that have to be made
up by additional ingredients. The latter usually turn spectator-model
operators into effective many-body
operators~\cite{Polyzou:1986df,Lev:1995wy}.

The spectator-model approach in point-form has been found to be specific,
since it preserves its spectator-model character in all reference
frames~\cite{Lev:1995wy}. This property is connected with the fact that
the generators of Lorentz transformations form the kinematic subgroup.
As a result the PFSM can be made manifestly covariant. Certainly, this
is a welcome behaviour for treating hadrons as relativistic few-quark
systems.

The PFSM has been found to produce surprisingly good results, especially
with regard to the elastic nucleon electroweak structure: the direct 
predictions by relativistic constituent quark models (CQM) have led
to quite a consistent description of all the relevant observables in 
remarkable overall agreement with existing experimental data at low
momentum transfers. In the PFSM relativistic (boost) effects definitely
have big influences in all respects. A similar situation has
been observed in 
the decay studies~\cite{Melde:2002ga,Melde:2004xj}. The covariant
predictions differ drastically from 
previous nonrelativistic or relativized 
results~\cite{Stancu:1989iu,Capstick:1993th,Glozman:1998xs,Theussl:2000sj}.
For the decays, however, the relativistic results systematically 
underestimate the experimental data and no satisfactory description 
is reached yet (with the simplistic decay models applied so far).
The characteristics of the results found in the
point-form approach have been seen in quite a similar manner with the 
relativistic CQM by the Bonn group~\cite{Loring:2001kx,Loring:2001ky}
in the framework of the Bethe-Salpeter
equation~\cite{Merten:2002nz,Metsch:2003ix,Metsch:2004qk}.

The detailed properties of the point-form approach using PFSM 
operators have not yet been fully understood. There have been several 
studies to elucidate its behaviour, also in comparison to the 
spectator approximation of other forms of 
RQM~\cite{Desplanques:2001zw,Desplanques:2001ze,Theussl:2003fs,Coester:2003rw}.
Concrete calculations with realistic CQM wave functions comparing the 
point and instant forms in completely analogous spectator-model
approaches have revealed big differences between
them~\cite{Berger:2005,Plessas:2004fn}. Of course, the different forms of RQM
are completely equivalent in a full calculation. Here, the question arises 
which contributions are effectively covered by the respective 
spectator-model approaches.

In this article, we deal with the fundamentals of defining PFSM operators.
In particular, we investigate the implications of the basic symmetries of 
the Poincar\'e group for their construction. The problem is studied 
for general (elastic or inelastic) transitions. It is made clear that the
PFSM in fact represents an effective many-body transition operator.
Implications of different ways of spectator-model constructions in point
form are demonstrated along concrete calculations of pionic decay widths of   
$N$ and $\Delta$ resonances for the case of the 
Goldstone-boson-exchange (GBE) CQM~\cite{Glozman:1998ag,Glozman:1998fs}.

\section{PFSM operators}
\label{sec:pfsm}

The general translational-invariant amplitude between certain
incoming and outgoing baryon states, $\left|V,M,J,\Sigma\right>$
and $\left<V',M',J',\Sigma'\right|$, is given by
\begin{multline}
\left<V',M',J',\Sigma'\right|{\hat O}
\left|V,M,J,\Sigma\right>\\
=\left<V',M',J',\Sigma'\right|{\hat O}_{\rm rd}
\left|V,M,J,\Sigma\right> \\
2MV_0\delta^3\left(M{\vec V}-M'{\vec V}'-{\vec Q}\right) \, ,
\label{operator}
\end{multline}
where $\hat O$ represents any electromagnetic, weak, or hadronic 
operator, and $\hat O_{\rm rd}$ is its reduced part. The baryon 
states are eigenstates of the four-velocity operator $\hat V$, the 
interacting mass operator $\hat M$, the (total) spin operator
$\hat J$, and its z-component $\hat \Sigma$ (the corresponding letters 
without a hat denoting their eigenvalues). The factor in front of 
the $\delta$-function is the invariant measure ensuring the 
correct normalization and transformation properties of the states. The 
$\delta$-function itself expresses the overall momentum conservation 
of the transition amplitude under the four-momentum transfer
$Q^{\mu}={P}^{\mu}-{P'}^{\mu}$ (for on-shell particles).
With the appropriate basis representations of the baryon eigenstates 
(see the Appendix) and inclusion of the necessary Lorentz transformations
the expression for the transition amplitude becomes
\begin{eqnarray}
&&\left<V',M',J',\Sigma'\right|{\hat O}
\left|V,M,J,\Sigma\right>=
\nonumber\\
&&
\frac{2}{MM'}\sum_{\sigma_i\sigma'_i}\sum_{\mu_i\mu'_i}{
\int{
d^3{\vec k}_2d^3{\vec k}_3d^3{\vec k}'_2d^3{\vec k}'_3
}}
\nonumber\\
&&
\sqrt{\frac{\left(\omega_1+\omega_2+\omega_3\right)^3}
{2\omega_1 2\omega_2 2\omega_3}}
\sqrt{\frac{\left(\omega'_1+\omega'_2+\omega'_3\right)^3}
{2\omega'_1 2\omega'_2 2\omega'_3}}
\nonumber\\
&&
\Psi^\star_{M'J'\Sigma'}\left({\vec k}'_1,{\vec k}'_2,{\vec k}'_3;
\mu'_1,\mu'_2,\mu'_3\right) \nonumber \\
&&
\prod_{\sigma'_i}{D_{\sigma'_i\mu'_i}^{\star \frac{1}{2}}
\left\{R_W\left[k'_i;B\left(V'\right)\right]\right\}
}
\nonumber\\
&&
\left<p'_1,p'_2,p'_3;\sigma'_1,\sigma'_2,\sigma'_3\right|{\hat O}_{\rm rd}
\left|p_1,p_2,p_3;\sigma_1,\sigma_2,\sigma_3\right>
\nonumber\\
&&
\prod_{\sigma_i}{D_{\sigma_i\mu_i}^{\frac{1}{2}}
\left\{R_W\left[k_i;B\left(V\right)\right]\right\}
} \nonumber \\
&&
\Psi_{MJ\Sigma}\left({\vec k}_1,{\vec k}_2,{\vec k}_3;\mu_1,\mu_2,\mu_3\right)
\nonumber\\
&&
2MV_0\delta^3\left(M{\vec V}-M'{\vec V}'-{\vec Q}\right)\, .
\label{transampl}
\end{eqnarray}
The integral measures stem from the completeness relation of the 
velocity states (see eq.~(\ref{eq:velcomp})), where the integrations over the 
velocities have already been performed exploiting the 
$\delta$-functions in the velocity-state representations of the baryon 
states (eq. (\ref{eq:wavefunc})). In this formula the individual quark momenta
$\vec k_i$ (and similarly ${\vec k}'_i$) are restricted by the
rest-frame condition $\sum_i{{\vec k}_i}=\vec 0$. The Wigner 
rotations stem from the Lorentz transformations to the boosted 
incoming and outgoing states, which have nonzero total momenta
$\vec P=M{\vec V}$ and $\vec P'=M'{\vec V}'$, respectively.  
The wave functions $\Psi^{\star}_{M'J'\Sigma'}$ and
$\Psi_{MJ\Sigma}$ denote the (rest-frame) velocity-state
representations of the baryon states. 
The reduced operator ${\hat O}_{\rm rd}$ remains sandwiched between
the free three-quark states.

At the outset ${\hat O}_{\rm rd}$ represents a general many-body operator. 
With present means the complete transition amplitude cannot be
computed for any one of the reactions in question (electromagnetic, weak,
or hadronic). Rather one has to resort to simplifications. 
Usually one first adopts a spectator model where the external particle 
couples only to one of the constituent quarks, while the other two are 
treated as spectators. In a nonrelativistic framework this would lead
to a genuine one-body operator. However, this is not the case in a 
Poincar\'e-invariant theory. Observing all necessary constraints one
arrives at effective many-body operators involving all quarks. This is
basically true in all forms of
RQM~\cite{Polyzou:1986df,Sengbusch:2004sf,Lev:1995wy,Simula:2001wx}.
Here we shall discuss the pertinent aspects especially for the point
form.

The Graz group has applied spectator model operators in the point 
form to several processes. In case of electromagnetic reactions the
PFSM for the current operator ${\hat J}^{\mu}_{\rm rd}$
reads~\cite{Klink:1998pr,Wagenbrunn:2000es,Boffi:2001zb}
\begin{multline}
\left<p'_1,p'_2,p'_3;\sigma'_1,\sigma'_2,\sigma'_3\right|
{\hat J}^{\mu}_{\rm rd}
\left|p_1,p_2,p_3;\sigma_1,\sigma_2,\sigma_3\right>
=\\
3 {\cal N}
\left<p'_1,\sigma'_1\right|\hat{J}^{\mu}_{\rm spec}\left|p_1,\sigma_1\right> \\
2p_{20}\delta\left({\vec p}_2-{\vec p}'_2\right)
2p_{30}\delta\left({\vec p}_3-{\vec p}'_3\right)
 \delta_{\sigma_{2}\sigma'_{2}}
   \delta_{\sigma_{3}\sigma'_{3}}
\label{eq:emcurr1}\, ,
\end{multline}
where the matrix element of the spectator-quark current has a formal
single-particle structure
\begin{multline}
\left<p'_1,\sigma'_1\right|\hat{J}^{\mu}_{\rm spec}\left|p_1,\sigma_1\right>= \\ 
e_1{\bar u}\left(p'_1,\sigma'_1\right)
\left[f_1({\tilde Q}^2) \gamma^\mu
+\frac{i}{2m_1} f_2({\tilde Q}^2)\sigma^{\mu\nu}
{\tilde q}_\nu\right]
u\left(p_1,\sigma_1\right) \,
\label{eq:emcurr2}
\end{multline}
with $f_1$ and $f_2$ being the Dirac form factors of the struck quark
with mass $m_1$. Its
spinor is expressed in terms of the usual two-component Pauli 
spinor $\chi$ in the following way
\begin{equation}
u(p,\sigma)= \sqrt{p_0+m}\left(\begin{array}{c}
\chi \\ \\
\frac{\vec{\sigma}\cdot\vec{p}}{p_0+m}\,\,\chi
\end{array}\right) \, .
\end{equation}

The PFSM for the axial current is defined in an analogous 
manner~\cite{Glozman:2001zc,Boffi:2001zb}
\begin{multline}
\left<p'_1,p'_2,p'_3;\sigma'_1,\sigma'_2,\sigma'_3\right|
{\hat A}^{\mu}_{a,\rm rd}
\left|p_1,p_2,p_3;\sigma_1,\sigma_2,\sigma_3\right>
=\\
3 {\cal N}
\left<p'_1,\sigma'_1\right|\hat{A}^{\mu}_{a,{\rm spec}}
\left|p_1,\sigma_1\right> \\
2p_{20}\delta\left({\vec p}_2-{\vec p}'_2\right)
2p_{30}\delta\left({\vec p}_3-{\vec p}'_3\right)
 \delta_{\sigma_{2}\sigma'_{2}}
   \delta_{\sigma_{3}\sigma'_{3}}
\label{eq:axcurr1}\, ,
\end{multline}
where the matrix element of the spectator-quark axial current is taken as
\begin{multline}
\left<p'_1,\sigma'_1\right|\hat{A}^{\mu}_{a,{\rm spec}}
\left|p_1,\sigma_1\right>= \\ 
{\bar u}\left(p'_1,\sigma'_1\right)
\left[g_A^q \gamma^\mu 
+ \frac{2f_\pi}{\widetilde{Q}^2+m_\pi^2} g_{qq\pi}\widetilde{q}^\mu\right]
\gamma_5 \frac{1}{2}{\tau}_a u\left(p_1,\sigma_1\right) \,
\label{eq:axcurr2}
\end{multline}
with $m_\pi$ the pion mass, $f_\pi$ the pion decay constant,
$g_A^q=1$ the quark axial charge,
$g_{qq\pi}$ the pion-quark coupling constant, and ${\tau}_a$ the 
isospin matrix with Cartesian index $a$.

For the mesonic decays of baryon resonances a decay model has been 
assumed with an elementary pseudovector coupling of the meson being 
directly emitted from a single quark. The corresponding PFSM
operator has the following
structure~\cite{Melde:2002ga,Melde:2004ce}
\begin{multline}
\left<p'_1,p'_2,p'_3;\sigma'_1,\sigma'_2,\sigma'_3\right|
{\hat D}_{m,{\rm rd}}\left|p_1,p_2,p_3;\sigma_1,\sigma_2,\sigma_3\right>
=\\
3{\cal N}
\frac{i g_{qqm}}{2m_1\left(2\pi\right)^{\frac{3}{2}}}
{\bar u}\left(p'_1,\sigma'_1\right)
\gamma_5\gamma^\mu \lambda_m
u\left(p_1,\sigma_1\right)
Q_\mu
\\
2p_{20}\delta\left({\vec p}_2-{\vec p}'_2\right)
2p_{30}\delta\left({\vec p}_3-{\vec p}'_3\right)
 \delta_{\sigma_{2}\sigma'_{2}}
   \delta_{\sigma_{3}\sigma'_{3}}
\label{eq:hadrcurr}\, 
\end{multline}
with the flavor matrix $\lambda_m$ characterizing the particular
decay mode and $g_{qqm}$ representing the corresponding quark-meson 
coupling constant.  

In eqs.~(\ref{eq:emcurr2}) and (\ref{eq:axcurr2}) ${\tilde q}^{\mu}$ 
denotes the momentum transfer to the struck quark in the Breit frame
\begin{equation}
{\tilde q}^{\mu}={p}^{\mu}_1-{p'}^{\mu}_1\,\,\,\,\, , \,\,\,\,\,
{\tilde q}^{\mu}{\tilde q}_{\mu}=-{\tilde Q^2}\, .
\end{equation}
It is different from the momentum $Q^{\mu}$ transferred to the baryon 
as a whole. Only part of the total momentum is transferred
to the struck quark. Even though the 
external particle (the photon or the intermediate boson) couples only 
to a single quark, also the spectator
quarks participate in the process since the total-momentum operator 
$\hat P$ is dynamical. 
This makes the PFSM current an effective many-body current. The same
consideration holds for the decay process when the meson is emitted 
from a single quark. In all cases ${\tilde q}^{\mu}$ is completely
fixed by the two spectator conditions and the overall momentum
conservation, and there is no arbitrariness.

The PFSM currents in eqs.~(\ref{eq:emcurr2}) and (\ref{eq:axcurr2})
maintain their spectator-model 
character in all reference frames~\cite{Lev:1995wy}. This is simply a 
consequence of the fact that the generators of Lorentz 
transformations are kinematical in the point form. The constructions 
of the electroweak currents and likewise of the decay operator in 
eq.~(\ref{eq:hadrcurr}) themselves are Lorentz-invariant and 
the spectator conditions are given by invariant $\delta$-functions.
The Lorentz-covariance of the spectator-model operators does not 
exist in the other forms of RQM. It is a specific property of the 
point form~\cite{Coester1995ic}. 

In eqs.~(\ref{eq:emcurr1}), (\ref{eq:axcurr1}), and (\ref{eq:hadrcurr})
there occurs a normalization factor $\cal N$ for all PFSM operators. 
In the electromagnetic case it is needed to 
reproduce the proton charge (the electric form factor at zero 
momentum transfer). For consistency reasons the same construction
should be adopted also for the weak and hadronic operators. The Graz 
group made the choice
\begin{eqnarray}
{\cal N}={\cal N}_{\rm S}&=&
\left(\frac{M}{\sum_i{\omega_i}}
\frac{M'}{\sum_i{\omega'_i}}\right)^{\frac{3}{2}}
\, .
\label{eq:nsymm}
\end{eqnarray}
In this way, $\cal N$ is assumed in a Lorentz-invariant form. This 
maintains the manifest covariance of the point-form transition 
amplitudes in the spectator model. From eq.~(\ref{eq:nsymm}) it is 
also seen that the normalization factor depends on the interactions, 
since $M$ and $M'$ are the eigenvalues of the interacting mass operator
(of the incoming and outgoing baryons). The 
ratios of the interacting mass eigenvalue to the sum of the 
individual quark energies in the denominator are chosen in a 
symmetric manner for both the incoming and outgoing channels. It
should be noted that in the integration of the matrix element in 
eq.~(\ref{transampl}) the sums of the individual quark energies in
$\cal N$ enter as functions of the integration variables and thus
indirectly introduce a certain dependence on the momentum transfer
$Q$ in the final results.

With the use of eq.~(\ref{eq:nsymm}) for $\cal N$, accounting for charge
normalization in a Poincar\'e-invariant PFSM, one achieved a very
reasonable description of the elastic electroweak nucleon structure
with both the relativistic Goldstone-boson-ex\-change and 
one-gluon-exchange CQMs~\cite{Plessas:2004fn}. In 
particular, the momentum dependence of the electromagnetic and axial 
form factors could be well reproduced by the PFSM for momentum
transfers up to $Q^2 \sim 5$ GeV$^2$.

The choice of the normalization factor $\cal N$, however, is not 
unique at this point. Under the assumptions of Poincar\'e invariance 
and charge normalization several other possibilities exist. This has 
to be considered especially in the context of inelastic processes.
Here one could also think of choices other than the symmetric one.
In the following we investigate the origin of the normalization
factor and study the implications of alternative forms in case of
the mesonic decays of $N$ and $\Delta$ resonances.

\section{Spectator Conditions and Translational Invariance}
\label{sec:implications}

In order to get a better insight into the nature of the normalization
factor $\cal N$ let us now shed some more light on the interplay of 
the spectator $\delta$-functions and the overall-momentum conserving
$\delta$-function in the expression for the matrix element of
the transition amplitude 
in eq.~(\ref{transampl}). This investigation will also further 
elucidate the role of ${\tilde q}^{\mu}$ and the proportioning of the 
whole momentum transfer among the individual quarks.

In point form we have
\begin{equation}
{\hat P}_{\rm free}={\hat M}_{\rm free}{\hat V}_{\rm free}
\end{equation}
for the free system and
\begin{equation}
\hat P=\hat M{\hat V}_{\rm free}
\end{equation}
for the interacting system according to the Bakamjian-Thomas 
construction~\cite{Bakamjian:1953}. The four-velocity remains 
kinematical upon introducing the interactions.
As a consequence the momentum and mass 
eigenvalues of the free and interacting systems are constrained by
\begin{equation}
V_{\rm free}=\frac{P_{\rm free}}{M_{\rm free}}=
\frac{\sum_i{p_i}}{\sum_i{\omega_i}}=
\frac{P}{M}=V\, .
\label{eq:constr}
\end{equation}
However, 
\begin{equation}
\sum_i{p_i}\ne P=MV \, .
\end{equation}
Remember that in point form the mass and the four-momentum operators
are the only operators affected by interactions. All other generators
of the Poincar\'e algebra remain kinematical.

Next we exploit the relation (\ref{eq:constr}) for the
overall-momen\-tum conserving $\delta$-function. 
Since it is an invariant form, 
we can consider it in any reference frame. Let us assume we are
working in the rest frame of the incoming baryon, where
$\sum_{i}{{\vec p}_{i,\rm in}}={\vec 0}$. For this particular case we 
denote all frame-dependent quantities with the index `$\rm in$'. Then
we can write the invariant $\delta$-function as
\begin{eqnarray}
&& 2MV_0\delta^3\left(M{\vec V}-M'{\vec V}'-{\vec Q}\right)
\nonumber \\
&&
=2MV_{0,\rm in}\delta^3\left(M{\vec V}_{\rm in}-M'{\vec V}'_{\rm in}
-{\vec Q}_{\rm in}\right)
\nonumber \\
&&
=2MV_{0,\rm in}\delta^3\left(-M'
\frac{\sum_i{{\vec p}'_{i,\rm in}}}{\sum_i{\omega'_i}}
-{\vec Q}_{\rm in}\right)
\\
&&
=2MV_{0,\rm in}\left(\frac{\sum_i{\omega'_i}}{M'}\right)^3
 \delta^3\left(-\sum_i{{\vec p}'_{i,\rm in}}
-\frac{\sum_i{\omega'_i}}{M'}{\vec Q}_{\rm in}\right)\nonumber \, .
\end{eqnarray}
Utilizing the spectator conditions,
$\vec p'_{2,\rm in}=\vec p_{2,\rm in}$ and
$\vec p'_{3,\rm in}=\vec p_{3,\rm in}$, this finally leads to the result
\begin{multline}
\label{eq:factorin}
2MV_{0,\rm in}\delta^3\left(M{\vec V}_{\rm in}-M'{\vec V}'_{\rm in}
-{\vec Q}_{\rm in}\right) \\
=2MV_{0,\rm in}\left(\frac{\sum_i{\omega'_i}}{M'}\right)^3
 \delta^3\left({\vec p}_{1,\rm in}-{\vec p}'_{1,\rm in}
-\frac{\sum_i{\omega'_i}}{M'}{\vec Q}_{\rm in}
\right)\, ,
\end{multline}
where ${V}_{0,\rm in}=1$ in this particular frame.

It is immediately evident that the momentum transfer to the struck 
quark is not the same as the momentum transfer to the baryon as a 
whole:
\begin{equation}
\frac{\sum_i{\omega'_i}}{M'}{\vec Q}_{\rm in}=
{\tilde {\vec q}}_{\rm in}
\ne \vec Q_{\rm in} \, .
\end{equation}
The difference can be expressed as
\begin{equation}
\vec Q_{\rm in} - {\tilde {\vec q}}_{\rm in}=
\frac{M'-\sum_i{\omega_i}'}{M'}{\vec Q}_{\rm in}=
\frac{M'-M'_{\rm free}}{M'}{\vec Q}_{\rm in}
\label{eq:3q}
\, .
\end{equation}
Obviously, it depends on the energies (and thus momenta) of all three 
quarks and therefore
makes the PFSM operator an effective many-body operator. 

It is also interesting to observe that the difference between the 
momentum transfers $\vec Q_{\rm in}$ and ${\tilde {\vec q}}_{\rm in}$ is 
determined by the difference between the interacting and the free
mass operators of the outgoing baryon, i.e. the interaction
responsible for its binding.

Now we notice the factor in front of the $\delta$-function in the last 
line of eq.~(\ref{eq:factorin}). Its nature resembles the 
inverse of the normalization factor $\cal N$ used in the definition 
of the PFSM operators in eq.~(\ref{eq:nsymm}). Only it is not symmetric 
in the mass eigenvalues of the incoming and outgoing baryons. However, 
it can be utilized for defining another normalization factor
\begin{equation}
{\cal N}_{\rm in}=\left(\frac{M'}{\sum_i{\omega'_i}}\right)^3
\, ,
\label{eq:nin}
\end{equation}
which would also work in the construction of the PFSM operators. In 
particular, it would guarantee for the correct proton charge
normalization and fulfill all the other constraints of Poincar\'e
invariance. Of course, the normalization factor ${\cal N}_{\rm in}$
produces a momentum dependence of the results different from the one
of ${\cal N}_{\rm S}$.

We can repeat the above procedure with the overall-momentum conserving
$\delta$-function in the rest-frame of
the outgoing baryon, where $\sum_{i}{{\vec p}'_{i,\rm out}}={\vec 0}$
and the index `$\rm out$' now denotes the frame-dependent variables in 
this specific frame.
We obtain instead of eq.~(\ref{eq:factorin}) the result
\begin{eqnarray}
\label{eq:factorout}
&&
2M{V}_0\delta^3\left(M{\vec V}-M'{\vec V}'-{\vec Q}\right) \nonumber \\
&&=2MV_{0,\rm out}\delta^3\left(M{\vec V}_{\rm out}-M'{\vec V}'_{\rm out}
-{\vec Q}_{\rm out}\right) \\
&&=2M{V}_{0,\rm out}\left(\frac{\sum_i{\omega_i}}{M}\right)^3
 \delta^3\left({\vec p}'_{1,\rm out}-{\vec p}_{1,\rm out}
-\frac{\sum_i{\omega_i}}{M}{\vec Q}_{\rm out}\right). \nonumber
\end{eqnarray}
Again, the momentum transfer to the struck quark is not the same
as the momentum transfer to the baryon as a whole
\begin{equation}
\frac{\sum_i{\omega_i}}{M}{{\vec Q}_{\rm out}}={\tilde {\vec q}}_{\rm out}
\ne {\vec Q}_{\rm out} \, ,
\end{equation}
where their difference now depends on the masses in the incoming 
channel
\begin{equation}
{\vec Q}_{\rm out} - {\tilde {\vec q}}_{\rm out}=
\frac{M-\sum_i{\omega_i}}{M}{{\vec Q}_{\rm out}}=
\frac{M-M_{\rm free}}{M}{{\vec Q}_{\rm out}}
\label{eq:3qout} \, .
\end{equation}
We note that in general ${\tilde {\vec q}}_{\rm out}$ is different from
${\tilde {\vec q}}_{\rm in}$. The portion of momentum transfer to the 
struck quark changes with the reference frame. The final result for 
any transition amplitude, however, does not. It is covariant in point 
form.

The factor in front of the $\delta$-function in the last 
line of eq.~(\ref{eq:factorout}) suggests again another
normalization factor
\begin{equation}
{\cal N}_{\rm out}=\left(\frac{M}{\sum_i{\omega_i}}\right)^3
\, .
\label{eq:nout}
\end{equation}
It would similarly be suited in the construction of PFSM operators just
like ${\cal N}_{\rm S}$ and ${\cal N}_{\rm in}$.

At this point all three normalization factors ${\cal N}_{\rm S}$,
${\cal N}_{\rm in}$, and ${\cal N}_{\rm out}$ meet the requirements posed
so far, namely, Poincar\'e invariance and charge normalization. Therefore 
we have to notice an ambiguity. In the next section we shall 
investigate the implications of the different possible choices in the 
mesonic decays of $N$ and $\Delta$ resonances, where we have different 
particles in the incoming and outgoing channels.

\section{$\cal N$ Dependence of Decay Widths}
\label{sec:NumRes}

Let us assume for the normalization factor $\cal N$ the general form
\begin{equation}
{\cal N}\left(y\right)=
\left(\frac{M}{\sum_i{\omega_{i}}}\right)^{3y}
\left(\frac{M'}{\sum_i{\omega'_{i}}}\right)^{3\left(1-y\right)}\, .
\label{eq:offsymfac}
\end{equation}
with $0\le y \le 1$. It contains all of the three forms,
${\cal N}_{\rm S}$, ${\cal N}_{\rm in}$, and ${\cal N}_{\rm out}$,
discussed in the previous section and it also meets all the
requirements posed (Poincar\'e invariance and charge normalization).
We emphasize that the transition amplitudes are covariant for every 
particular fixed $y$. However, the results will vary for different
values of the asymmetry parameter $y$ .

We studied the decay widths of $N$ and $\Delta$ resonances with the 
PFSM decay operator of eq.~(\ref{eq:hadrcurr}). The actual 
computations were performed in the rest frame of the decaying resonance. 
We emphasize, however, that the PFSM results are frame-independent.
In table~\ref{tab1} 
we first demonstrate the dissimilarity of the predictions in case of
the GBE CQM for either one of the choices ${\cal N}_{\rm S}$,
${\cal N}_{\rm in}$, and ${\cal N}_{\rm out}$. It is seen that the 
normalization factor ${\cal N}_{\rm in}$ yields 
the smallest values of the decay widths in all cases. Contrary to 
that, ${\cal N}_{\rm out}$ always produces the biggest predictions. 
The symmetric ${\cal N}_{\rm S}$ is intermediate between the two.
In the last column of table~\ref{tab1} we have also quoted the 
results that would be obtained if $\cal N$ was left out completely; 
we denoted this case by ${\cal N}_{\rm bare}$. It does not meet the 
requirement of (proton) charge normalization in the electroweak sector.
Obviously, the corresponding results are completely unreasonable also 
here for the decay widths.

When considering the different predictions in table~\ref{tab1} we 
notice that the normalization factor ${\cal N}_{\rm out}$ leads to
an overestimation of the experimental data in several cases. This 
drawback is avoided completely by ${\cal N}_{\rm S}$. The 
corresponding results always remain smaller than the experimental 
data or do not exceed them. We consider this to be a reasonable
feature, since a more elaborate decay model than the one used here
would tentatively bring in additional contributions that are 
expected to make the predictions bigger and thus get them into closer
agreement with the data. The results with ${\cal N}_{\rm in}$ are 
always much too small as compared to experiment.

\begin{figure}
\resizebox{0.45\textwidth}{!}{%
  \includegraphics[clip=]
{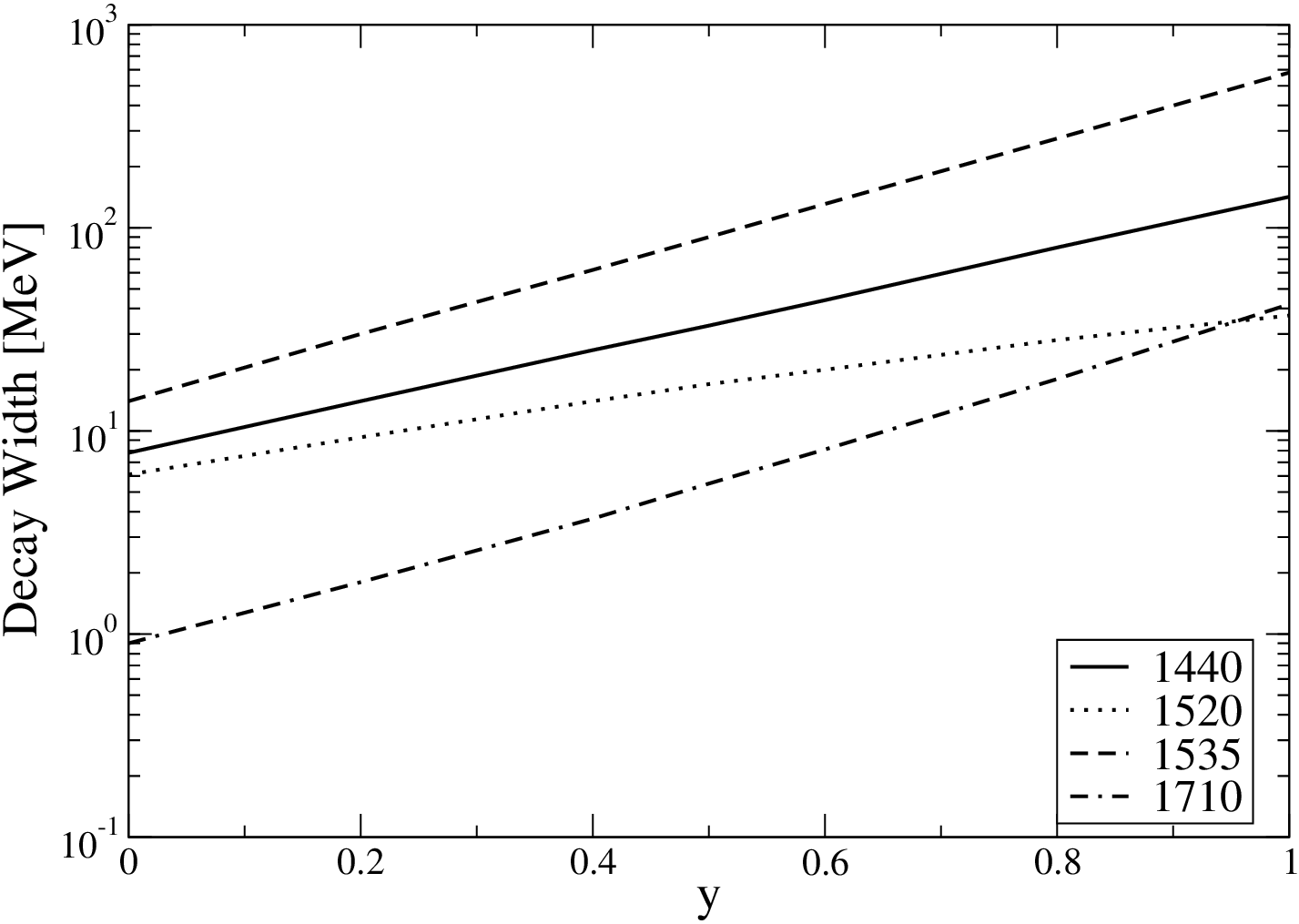}
}
\caption{
Dependence of the $\pi$ decay widths on the asymmetry parameter $y$ 
in the normalization factor of eq. (\ref{eq:offsymfac}) for selected
$N$ resonances.
}
\label{fig:offsym1}      
\end{figure}

\begin{figure}
\resizebox{0.45\textwidth}{!}{%
  \includegraphics[clip=]
{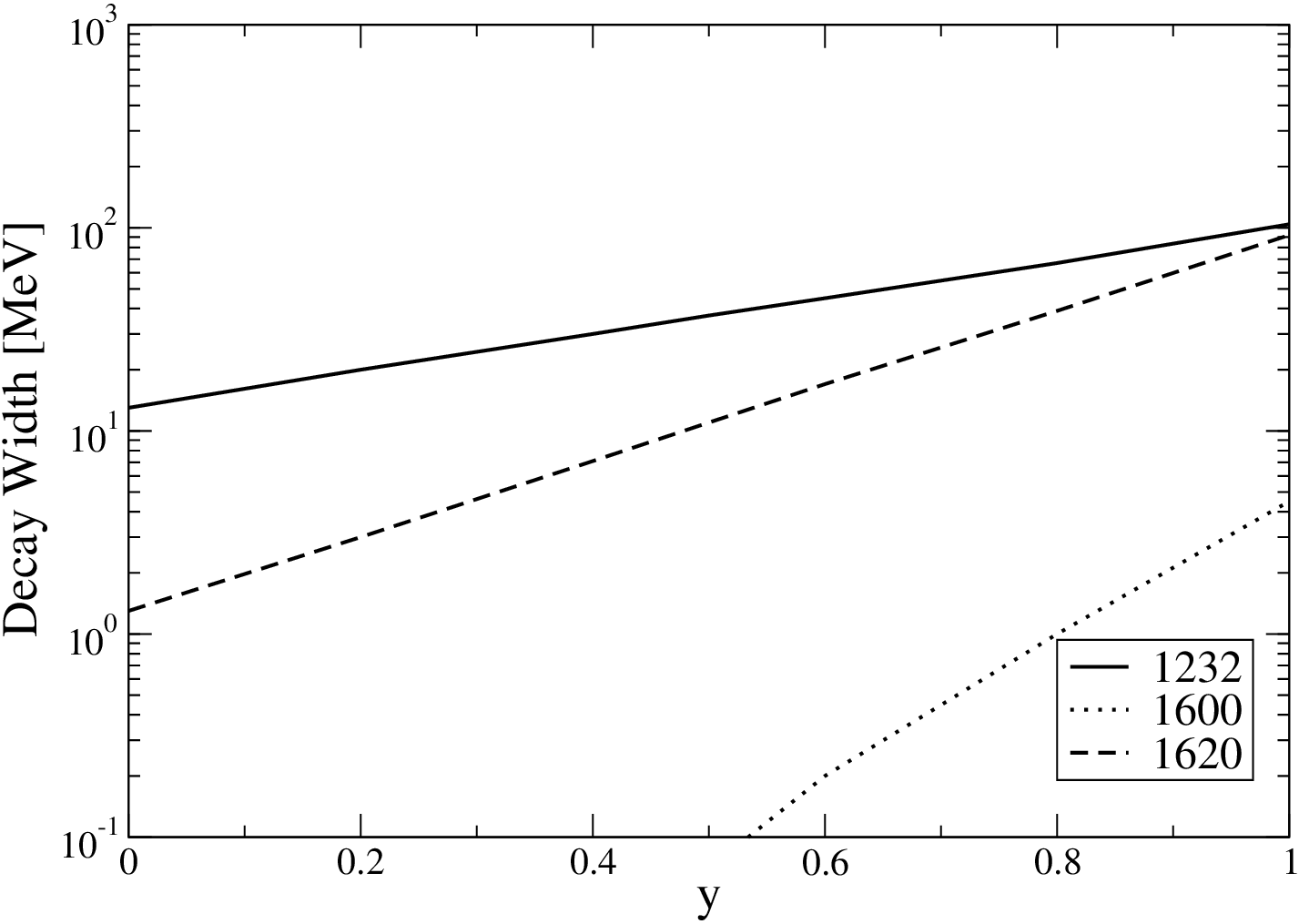}
}
\caption{
Dependence of the $\pi$ decay widths on the asymmetry parameter $y$
in the normalization factor of eq. (\ref{eq:offsymfac}) for selected
$\Delta$ resonances.
}
\label{fig:offsym2}      
\end{figure}

The detailed dependence of the results on the asymmetry parameter $y$
in eq.~(\ref{eq:offsymfac}) is demonstrated in table~\ref{tab2}. A smooth 
transition of the theoretical values from $y=0$ (corresponding to 
${\cal N}_{\rm in}$) via $y=0.5$ (corresponding to ${\cal N}_{\rm S}$) 
to $y=1$ (corresponding to ${\cal N}_{\rm out}$) is observed for all 
resonance decays. We have exemplified this behaviour for some of the 
$N$ and $\Delta$ resonances in figs.~\ref{fig:offsym1} and
\ref{fig:offsym2}, respectively.
There is practically an exponential rise of the theoretical values 
when $y$ varies from zero to one.

From these studies one learns that the normalization factors modify 
the dependence on the recoil, i.e. on the $Q$-dependence in 
eq.~(\ref{eq:hadrcurr}). It is noteworthy that the symmetric choice 
${\cal N}_{\rm S}$, which was originally adopted as the optimal one 
in the electroweak case, also leads to the most reasonable results in 
the hadronic decays considered here (in the sense that the 
experimental data are not overestimated).

\begin{figure}
\resizebox{0.45\textwidth}{!}{%
  \includegraphics[clip=]
{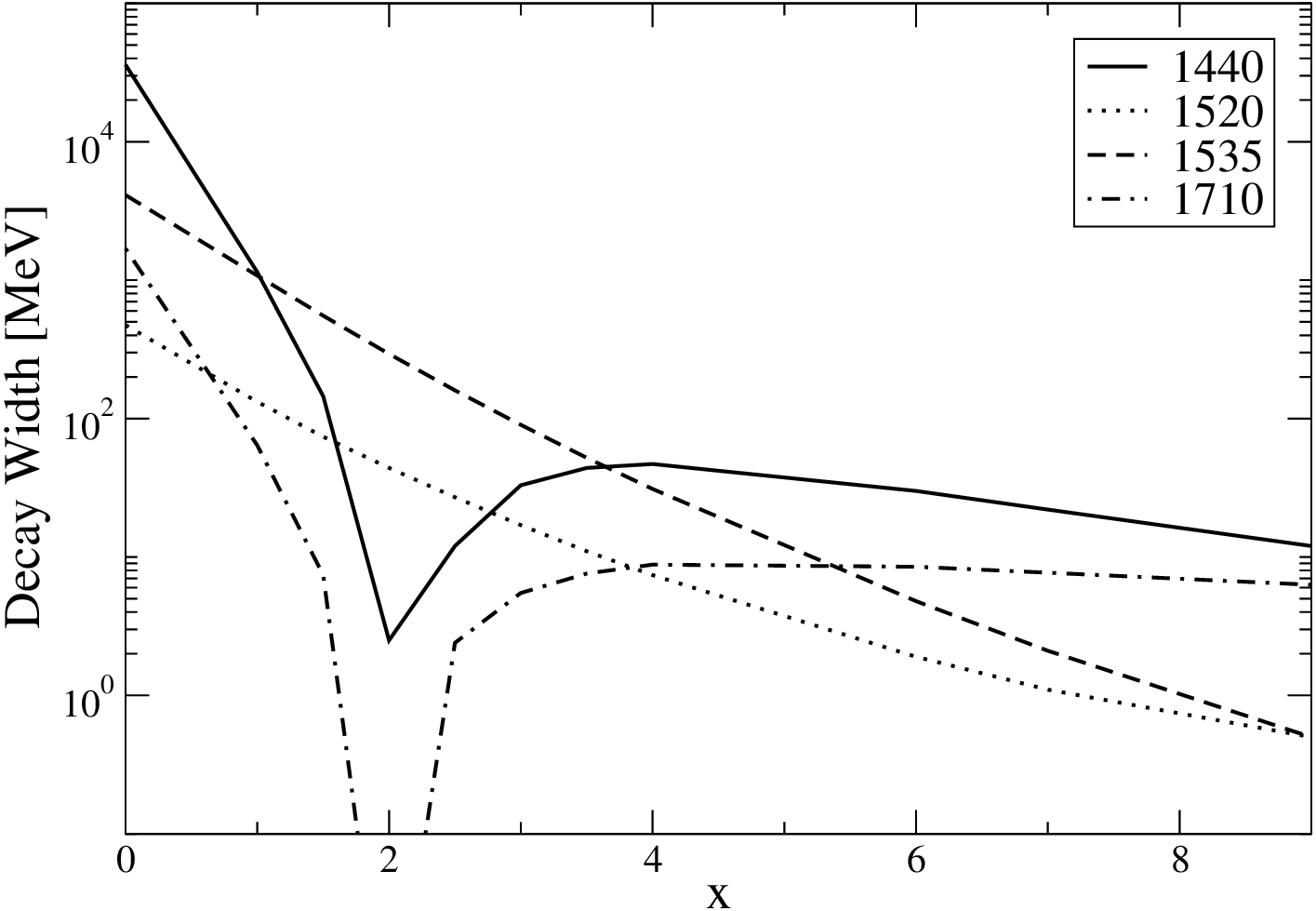}
}
\caption{Dependence of the $\pi$ decay widths on the exponent $x$ 
of the normalization factor in eq. (\ref{eq:exponent}) for selected
$N$ resonances.
}
\label{fig:Nexpo}      
\end{figure}

\begin{figure}
\resizebox{0.45\textwidth}{!}{%
  \includegraphics[clip=]
{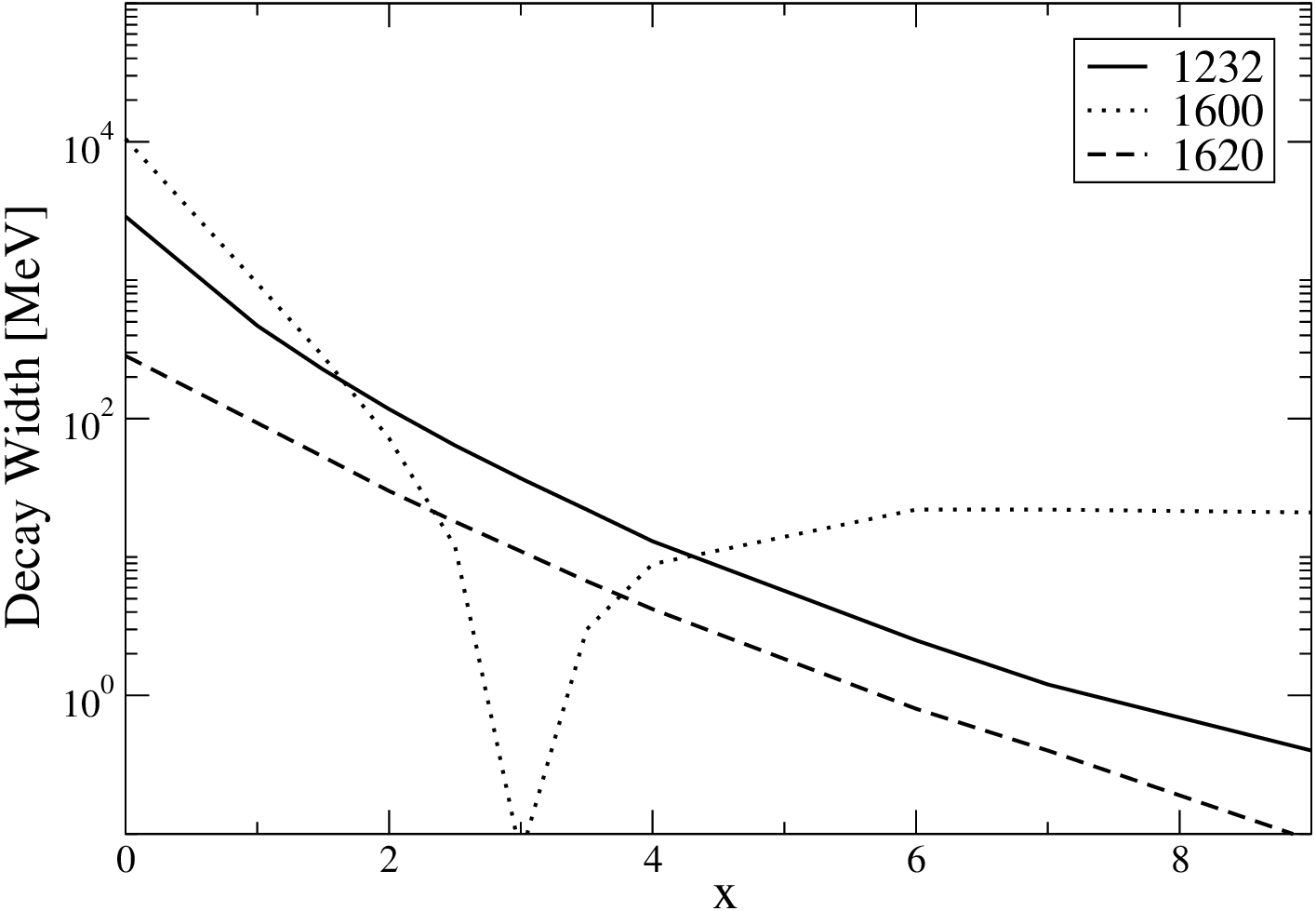}
}
\caption{
Dependence of the $\pi$ decay widths on the exponent $x$ 
of the normalization factor in eq. (\ref{eq:exponent}) for selected
$\Delta$ resonances.}
\label{fig:Dexpo}      
\end{figure}

That the normalization factor $\cal N$ in eqs.~(\ref{eq:emcurr1}),
(\ref{eq:axcurr1}), and (\ref{eq:hadrcurr})
effectively introduces a momentum cut-off can even better be seen if 
we investigate the form
\begin{equation}
\label{eq:exponent}
{\cal N}\left(x\right)
=
\left(\frac{M}{\sum_i{\omega_{i}}}\frac{M'}{\sum_i{\omega'_{i}}}\,
\right)^{\frac{x}{2}}\, 
\end{equation}
with an arbitrary exponent $x$. It still represents a 
Poincar\'e-invariant construction but it does not guarantee for the 
proper charge normalization unless $x=3$. It is instructive to look at 
the predictions for decay widths as a function of the exponent $x$
in table~\ref{tab:exponent}. Starting out from the (unreasonable) 
bare case the theoretical results evolve smoothly with increasing 
exponent $x$. For $x=3$ we recover the predictions for
${\cal N}_{\rm S}$ in table~\ref{tab1}. For certain resonances the 
decay widths have a minimum. This is exemplified in 
figs.~\ref{fig:Nexpo} and \ref{fig:Dexpo}. We notice that the minima 
occur just for the resonances $N(1440)$, $N(1710)$, and $\Delta(1600)$, 
which are known as the so-called structure-dependent 
resonances~\cite{Koniuk:1980vy}. They are the radial excitations of
the $N$ and $\Delta$ ground states, respectively, with a corresponding
nodal behaviour in their wave functions. These characteristics are
quite distinct from the other resonances, which show a monotonous 
dependence on the exponent $x$.

If we assume again the criterion that the theoretical predictions for 
decay widths with the decay operator~(\ref{eq:hadrcurr}) should not 
exceed the experimental data, we find as the optimal case the one 
with $x=3$. In this way we are led back to the assumption of the 
symmetric normalization factor in eq.~(\ref{eq:nsymm}), which also 
meets all the theoretical requirements imposed.

\section{Summary}

We have investigated the spectator-model construction of transition 
operators in point form. Based on the explicit forms of the current 
operators for electromagnetic as well as weak reactions
and the decay operator for baryon resonances
we have investigated the specific ingredients in the PFSM. 
We have made transparent the effective many-body nature of the 
spectator-model operator in point form. In particular, we have 
explained the distribution of the total momentum transfer $Q$ among 
the constituent quarks in a baryon through the Lorentz boosts
under the constraint of translational invariance. As a result the 
momentum $\tilde q$ transferred to the struck quark is only a part
of the total $Q$. Furthermore we have devoted our attention to the 
definition of the normalization factor $\cal N$ occurring in the PFSM 
operators. We noticed that several choices are possible even under 
the constraints of charge normalization (in the electroweak case) and 
Poincar\'e invariance. However, the final results depend on the
particular choice made. Therefore in any PFSM calculation it should
be specified which normalization factor has been adopted.

The influence of different normalization factors has been investigated 
with regard to pionic decays of $N$ and $\Delta$ resonances. In the 
concrete calculations we have employed the wave functions as produced 
by the GBE CQM. Qualitatively the results would be quite similar with 
other realistic baryon wave functions, e.g., the ones from a 
one-gluon-exchange CQM (see ref.~\cite{Melde:2004xj},
where the influences of different dynamics on decay widths have been
discussed). We have demonstrated the modification of the 
momentum dependence that is introduced by different choices of
$\cal N$. Upon comparing the theoretical predictions with experiment 
we have found a preference for the symmetric choice ${\cal N}_{\rm S}$ 
in the PFSM decay operator adopted here. This observation is congruent 
with the one made in the electroweak sector. With regard to the pionic
decays, however, even the calculation with ${\cal N}_{\rm S}$ using the
wave functions of the GBE CQM and the present PFSM decay operator does
not produce a satisfactory description of the experimental data.

While our investigations have been made specifically for the point 
form, most of the questions addressed here are also relevant for the 
other forms of RQM in case spectator operators are considered. Similar 
problems occur in the instant and front forms if the necessary 
invariance constraints are imposed (cf., for instance, refs.
\cite{Polyzou:1986df,Lev:1995wy,Simula:2001wx,Keister:1994mg,Lev:2000vm,%
Desplanques:2004zp}).
In some respects even additional 
complications arise connected with the fact that Lorentz 
transformations are no longer kinematical. For example, in instant 
form the spectator-model character of any operator constructed in one 
frame will not be maintained under Lorentz transformations. In any 
other reference frame specific genuine many-body operators will be 
generated.

Here, and in previous works, we have seen certain advantages of the 
point-form approach in the treatment of relativistic few-body problems.
The PFSM construction definitely includes effective contributions from
many-body operators. Their sources are twofold. They stem from the
sharing of the total momentum transfer to the individual quarks and
from the necessary normalization factor $\cal N$, which involves the
interacting mass operators. The different possible choices of $\cal N$
constitute a quantitatively significant ambiguity. 
It is important to take these peculiarities of the PFSM into account
before going to include explicit many-body operators. 

%
\acknowledgement{We should like to thank W. Klink as well as L. Glozman
and W. Schweiger for many useful discussions and B. Sengl for a 
careful reading of the manuscript.
This work was supported by the Austrian Science Fund (Project 
P16945). T.M. would like to thank the 
INFN and the Physics Department of the University of Padova for their 
hospitality and MIUR-PRIN for financial support.}
%
%
%
 \renewcommand{\theequation}{A\arabic{equation}}
  \setcounter{equation}{0}  
\section*{Appendix:\\
States and Wave Functions in Point Form RQM
}
\label{sec:Appendix}
In RQM baryon states are expressed as eigenstates
$\left|P, J,\Sigma\right>$ of the four-momentum $\hat P$, total spin 
$\hat J$, and its $z$-component $\hat \Sigma$ (the letters without hat 
denoting the corresponding eigenvalues). Their covariant 
normalization is
\begin{equation}
\left<P', J',\Sigma'|P, J,\Sigma\right>=
2P_0\delta^3\left({\vec P}-{\vec P}'\right)\delta_{JJ'}
\delta_{\Sigma\Sigma'}\, .
\label{normP}
\end{equation}
The baryon states can equivalently be expressed by 
$\left|V, M,J,\Sigma\right>$, i.e. as eigenstates of the four-velocity 
operator $\hat V$, the (interacting) mass operator $\hat M$ as well 
as $\hat J$ and $\hat \Sigma$. For the eigenvalues of the four-vector
$V$ one always has the constraint $V^\mu V_\mu=1$. The covariant
normalization of the baryon states $\left|V, M,J,\Sigma\right>$ reads
\begin{multline}
\left<V', M',J',\Sigma'|V, M,J,\Sigma\right>
\\
=
2MV_0\delta^3\left(M{\vec V}-M'{\vec V}'\right)\delta_{JJ'}
\delta_{\Sigma\Sigma'}\, .
\label{normV}
\end{multline}

For the baryon wave functions the eigenstates can be expressed in
different basis representations. One basis is provided by the free 
three-particle states, which are tensor products of one-particle 
states. Their covariant normalization is
\begin{multline}
\left<p'_1,p'_2,p'_3;\sigma'_1\sigma'_2,\sigma'_3|
p_1,p_2,p_3;\sigma_1\sigma_2,\sigma_3\right>
\\
=2p_{10}\delta^3\left({\vec p}_1-{\vec p}'_1\right)
2p_{20}\delta^3\left({\vec p}_2-{\vec p}'_2\right)
2p_{30}\delta^3\left({\vec p}_3-{\vec p}'_3\right)
\\
\times
\delta_{\sigma_1\sigma'_1}\delta_{\sigma_2\sigma'_2}
\delta_{\sigma_3\sigma'_3}\, ,
\label{eq:orthp}
\end{multline}
and their completeness relation reads
\begin{multline}
1=\sum_{\sigma_1,\sigma_2,\sigma_3}
\int{
\frac{d^3{\vec p}_1}{2p_{10}}\frac{d^3{\vec p}_2}{2p_{20}}
\frac{d^3{\vec p}_3}{2p_{30}}
}
\\
\left|p_1,p_2,p_3;\sigma_1,\sigma_2,\sigma_3\right>
\left<p_1,p_2,p_3;\sigma_1,\sigma_2,\sigma_3\right|
\, .
\label{eq:pcomp}
\end{multline}
For on-shell particles out of the 12 momentum components only 9 remain
as integration variables. In the rest frame we write these states as
$\left|k_1,k_2,k_3;\mu_1,\mu_2,\mu_3\right>$, where
$\sum_i{{\vec k}_i}=\vec 0$.

Another basis is provided by the so-called velocity states (of the
free system). They are advantageous in practical caluclations and 
have already been used before, among others in refs. 
\cite{Klink:1998hc,Krassnigg:2003gh}, with 
slightly different normalizations. We define them by
\begin{eqnarray}
&&\left|v;\vec{k}_1,\vec{k}_2,\vec{k}_3;\mu_1,\mu_2,\mu_3\right\rangle
=U_{B(v)}
\left|k_1,k_2,k_3;\mu_1,\mu_2,\mu_3\right\rangle
\nonumber \\
&&=\sum_{\sigma_1,\sigma_2,\sigma_3}
\prod\limits_{i=1}^3D^{\frac{1}{2}}_{\sigma_i\mu_i}[R_W(k_i,B(v))]
\left|p_1,p_2,p_3;\sigma_1,\sigma_2,\sigma_3\right\rangle \, .
\nonumber \\ 
\label{eq:velstates}
\end{eqnarray}
Here $B\left(v\right)$, with unitary representation 
$U_{B\left(v\right)}$,
is a boost with four-velocity $v$ on the three-body states
$\left|k_1,k_2,k_3;\mu_1,\mu_2,\mu_3\right\rangle$ in the rest frame.
The relation between $p_i$ and $k_i$ is thus given by
\begin{equation}
p_i=B\left(v\right)k_i\, ,
\end{equation}
and the four-velocity is expressed by
\begin{equation}
{v}=\frac{\sum_i{{p}_i}}{M_{\rm free}}
=\frac{\sum_i{{p}_i}}{\sum_i{\omega_i}}
\, ,
\end{equation}
where $M_{\rm free}$ is the invariant free mass and
$\omega_i=\sqrt{{\vec k}_i^2+m_i^2}$ are the energies
of the individual quarks with mass $m_i$.
In case of the velocity states
one has $\vec v$ and two of the three quark momenta,
$\vec k_2$ and $\vec k_3$, say, as the 9 independent variables. 
The transformation from the free three-body states to the free 
velocity states is given by the Jacobi determinant
\begin{equation}
{\cal J}\left\{\frac{
\partial \left({\vec p}_1,{\vec p}_2,{\vec p}_3\right)}
{\partial \left({\vec v},{\vec k}_2,{\vec k}_3\right)}
\right\}
=\frac{2p_{10}2p_{20}2p_{30}\left(\omega_1+\omega_2+\omega_3\right)^3}
{2\omega_1 2\omega_2 2\omega_3v_0}
\, .
\end{equation}
Due to eqs. (\ref{eq:orthp}) and (\ref{eq:pcomp}) it implies the
following normalization
\begin{multline}
    \left\langle v;\vec{k}_1,\vec{k}_2,\vec{k}_3;\mu_1,\mu_2,\mu_3
    |
    v';\vec{k}'_1,\vec{k}'_2,\vec{k}'_3;
\mu'_1,\mu'_2,\mu'_3\right\rangle\\
   =\frac{2\omega_{1}2\omega_{2}2\omega_{3}}
    {\left(\omega_{1}+\omega_{2}+\omega_{3}\right)^{3}}
     \delta_{\mu_{1}\mu'_{1}}\delta_{\mu_{2}\mu'_{2}}
\delta_{\mu_{3}\mu'_{3}}\\
     \times
     v_{0}\delta^{3}\left(\vec{v}-\vec{v}'\right)
    \delta^{3}\left(\vec{k}_{2}-\vec{k}'_{2}\right)
     \delta^{3}\left(\vec{k}_{3}-\vec{k}'_{3}\right)
\label{normvel}
\end{multline}
and completeness relation
\begin{multline}
{\mathbf 1}=\sum_{\mu_{1}\mu_{2}\mu_{3}}\int{
    \frac{d^{3}{\vec v}}{v_{0}}
    \frac{d^{3}{\vec k}_{2}}{2\omega_{2}}
    \frac{d^{3}{\vec k}_{3}}{2\omega_{3}}
    \frac{\left(\omega_{1}+\omega_{2}+
\omega_{3}\right)^{3}}{2\omega_{1}}
    }\\
    \times \left|v;\vec{k}_1,\vec{k}_2,\vec{k}_3;
\mu_1,\mu_2,\mu_3\right\rangle
    \left\langle 
    v;\vec{k}_1,\vec{k}_2,\vec{k}_3;\mu_1,\mu_2,\mu_3\right|
\label{eq:velcomp}
\end{multline}
for the velocity states.

The baryon wave function in any reference frame is given by the 
velocity-state representation of the eigenstates
$\left|V, M,J,\Sigma\right>$
\begin{multline}
    \left\langle v;\vec{k}_1,\vec{k}_2,\vec{k}_3;\mu_1,\mu_2,\mu_3
    |
    V,M,J,\Sigma\right\rangle
    =\frac{\sqrt{2}}{M} v_{0}\delta^{3}\left(\vec{v}-\vec{V}\right)\\ 
   \sqrt{\frac{2\omega_{1}2\omega_{2}2\omega_{3}}
    {\left(\omega_{1}+\omega_{2}+\omega_{3}\right)^{3}}
    }
   \Psi_{MJ\Sigma}\left(
   \vec{k}_1,\vec{k}_2,\vec{k}_3;\mu_1,\mu_2,\mu_3
   \right) \, .
   \label{eq:wavefunc}
\end{multline}
The wave functions
$\Psi_{MJ\Sigma}\left(\vec{k}_1,\vec{k}_2,\vec{k}_3;\mu_1,\mu_2,\mu_3\right)$
are normalized to unity
\begin{multline}
    \sum_{\mu_{1}\mu_{2}\mu_{3}}
    \int{d^{3}{\vec k}_{2}d^{3}{\vec k}_{3}} \\
    \times
   \Psi^{\star}_{M'J'\Sigma'}\left(
   \vec{k}_1,\vec{k}_2,\vec{k}_3;\mu_1,\mu_2,\mu_3
   \right)
   \Psi_{MJ\Sigma}\left(
   \vec{k}_1,\vec{k}_2,\vec{k}_3;\mu_1,\mu_2,\mu_3
   \right) \\
   =\delta_{MM'}\delta_{JJ'}\delta_{\Sigma\Sigma'}
\end{multline}
in concordance with the normalization conditions~(\ref{normV}) and
(\ref{normvel}). The advantage of this velocity-state representation
is that the motion of the system as a whole can always be separated 
from the internal motion represented in $\Psi_{MJ\Sigma}$.

%
%
%
%
%
%
%
%

\newpage
.
\newpage
\renewcommand{\arraystretch}{1.5}
\begin{table}
\caption{
PFSM predictions of the GBE CQM \cite{Glozman:1998ag} for $\pi$ decay
widths with different choices of the normalization factor $\cal N$ in 
eq. (\ref{eq:hadrcurr}).
}
\label{tab1}       
\begin{tabular}{crcccc}
\hline\noalign{\smallskip}
Decay 
&Experiment~\cite{Eidelman:2004wy}
& ${\cal N}_{\rm in}$ 
& ${\cal N}_{\rm S}$ 
& ${\cal N}_{\rm out}$
& ${\cal N}_{\rm bare}$ 
 \\
\noalign{\smallskip}\hline\noalign{\smallskip}
{\small $N(1440)$}
	&  $\left(227\pm 18\right)_{-59}^{+70}$	
	&  $7.8$ 	
	&  $33$ 
 	&  $142$ 
         &  $36004$ 
 \\
{\small $N(1520)$}
	&  $\left(66\pm \phantom{0}6
                 \right)_{-\phantom{0}5}^{+\phantom{0}9}$  
	&  $6.1$ 
	&  $17$ 
	&  $37$ 
	&  $474$ 
 \\
{\small $N(1535)$}
	&  $ \left(67\pm 15\right)_{-17}^{+28}$ 
	&  $14$ 	
	&  $90$ 
	&  $581$ 
	&  $4123$ 
\\
{\small $N(1650)$}
	&  $ \left(109\pm 26 \right)_{-\phantom{0}3}^{+36}$ 
	&  $3.5$ 
	&  $29$ 
	&  $242$ 
	&  $1110$ 
\\
{\small $N(1675)$}
	&  $ \left(68\pm \phantom{0}8\right)_{-\phantom{0}4}^{+14}$ 
	&  $1.3$ 
	&  $5.4$ 
	&  $19$ 
	&  $81$ 
\\
{\small $N(1700)$}
	&  $ \left(10\pm \phantom{0}5\right)_{-\phantom{0}3}^{+\phantom{0}3}$ 
	&  $0.04$ 
	&  $0.8$ 
	&  $10$ 
	&  $30$ 
\\
{\small $N(1710)$}
	&  $\left(15\pm \phantom{0}5\right)_{-\phantom{0}5}^{+30}$  
	&  $0.9$ 
	&  $5.5$ 
	&  $42$ 
	&  $1692$ 
\\
{\small $\Delta(1232)$}
	&  $\left(119\pm \phantom{0}1 \right)_{-\phantom{0}5}^{+\phantom{0}5}$ 
	&  $13$ 
	&  $37$ 
	&  $104$ 
	&  $2883$ 
\\
{\small $\Delta(1600)$}
	&  $\left(61\pm 26\right)_{-10}^{+26}$ 
	&  $0.0003$ 
	&  $0.07$ 
	&  $4.5$ 
	&  $10468$ 
\\
{\small $\Delta1620)$}
	&  $ \left(38\pm \phantom{0}8\right)_{-\phantom{0}6}^{+\phantom{0}8}$ 
	&  $1.3$ 
	&  $11$ 
	&  $92$ 
	&  $284$ 
\\
{\small $\Delta(1700)$}
	&  $ \left(45\pm 15\right)_{-10}^{+20}$ 
	&  $0.7$ 
	&  $2.3$ 
	&  $5.4$ 
	&  $26$ 
\\
\noalign{\smallskip}\hline
\end{tabular}
\end{table}
%
%
%
\begin{table*}
\caption{
Dependence of the $\pi$ decay widths on the asymmetry parameter $y$ 
in the normalization factor of eq. (\ref{eq:offsymfac}).
}
\label{tab2}       
\begin{center}
\begin{tabular}{cccccccccccc}
\hline\noalign{\smallskip}
  \rule{0pt}{1.2cm}
  $y$
 &  \begin{rotate}{90}$N(1440)$\end{rotate}
 &  \begin{rotate}{90}$N(1520)$\end{rotate} 
 &  \begin{rotate}{90}$N(1535)$ \end{rotate} 
 &  \begin{rotate}{90}$N(1650)$ \end{rotate} 
 &  \begin{rotate}{90}$N(1675)$ \end{rotate} 
 &  \begin{rotate}{90}$N(1700)$  \end{rotate}
 &  \begin{rotate}{90}$N(1710)$  \end{rotate}
 &  \begin{rotate}{90}$\Delta (1232)$\end{rotate} 
 &  \begin{rotate}{90}$\Delta (1600)$ \end{rotate}
 &  \begin{rotate}{90}$\Delta (1620)$\end{rotate}  
 &  \begin{rotate}{90}$\Delta (1700)$\end{rotate} \\
\noalign{\smallskip}\hline
$0.0$ & $7.8$   & $6.1$ & $14$ & $3.5$ & $1.3$ & $0.04$ & $0.87$ 
& $13$   & $0.0003$ & $1.3$ & $0.72$\\
$0.2$ & $14$   & $9.3$ & $30$ & $8.1$ & $2.3$ & $0.14$ & $1.8$ 
& $20$   & $0.001$ & $3.0$ & $1.2$\\
$0.4$ & $25$   & $14$ & $62$ & $19$ & $4.1$ & $0.44$ & $3.7$ 
& $30$   & $0.02$ & $7.1$ & $1.9$\\
$0.5$ & $33$   & $17$ & $90$ & $29$ & $5.4$ & $0.75$ & $5.5$ 
& $37$   & $0.07$ & $11$ & $2.3$\\
$0.6$ & $44$   & $20$ & $131$ & $44$ & $7.0$ & $1.3$ & $8.1$ 
& $45$   & $0.19$ & $17$ & $2.9$\\
$0.8$ & $80$   & $28$ & $275$ & $102$ & $12$ & $3.7$ & $18$ 
& $67$   & $0.99$ & $39$ & $4.1$\\
$1.0$ & $142$   & $37$ & $581$ & $242$ & $19$ & $10$ & $42$ 
& $104$   & $4.5$ & $92$ & $5.4$\\
\hline
Exp. & $227$ & $66$ & $67$ & $109$ & $68$ & $10$ & $15$
 & $119$ & $61$ & $38$ & $45$ \\
\noalign{\smallskip}\hline
\end{tabular}
\end{center}
\end{table*}
\begin{table*}
\caption{Dependence of the $\pi$ decay widths on the exponent $x$ 
in the normalization factor of eq. (\ref{eq:exponent}).
}
\label{tab:exponent}      
\begin{center}
\begin{tabular}{cccccccccccc}
\hline\noalign{\smallskip}
  \rule{0pt}{1.2cm}
  $x$
 &  \begin{rotate}{90}$N(1440)$ \end{rotate}
 &  \begin{rotate}{90}$N(1520)$ \end{rotate} 
 &  \begin{rotate}{90}$N(1535)$ \end{rotate} 
 &  \begin{rotate}{90}$N(1650)$ \end{rotate} 
 &  \begin{rotate}{90}$N(1675)$ \end{rotate} 
 &  \begin{rotate}{90}$N(1700)$  \end{rotate}
 &  \begin{rotate}{90}$N(1710)$  \end{rotate}
 &  \begin{rotate}{90}$\Delta (1232)$ \end{rotate} 
 &  \begin{rotate}{90}$\Delta (1600)$  \end{rotate}
 &  \begin{rotate}{90}$\Delta (1620)$ \end{rotate}  
 &  \begin{rotate}{90}$\Delta (1700)$ \end{rotate} \\
\hline\noalign{\smallskip}
$0.0$ & $36004$   & $474$ & $4123$ & $1110$ & $81$ & $30$ & $1692$ 
& $2883$   & $10468$ & $284$ & $26$\\
$1.0$ & $1143$   & $132$ & $1078$ & $256$ & $28$ & $7.6$ & $64$ 
& $469$   & $946$& $93$ & $11$\\
$1.5$ & $145$   & $74$ & $554$ & $140$ & $18$ & $4.1$ & $7.4$ 
& $226$   & $282$ & $53$ & $6.9$\\
$2.0$ & $2.5$   & $44$ & $293$ & $80$ & $12$ & $2.3$ & $0.002$ 
& $117$   & $72$ & $30$ & $4.7$\\
$2.5$ & $12$   & $27$ & $160$ & $47$ & $7.8$ & $1.3$ & $2.4$ 
& $64$   & $12$ & $18$ & $3.3$\\
$3.0$ & $33$   & $17$ & $90$ & $29$ & $5.4$ & $0.75$ & $5.5$ 
& $37$   & $0.07$ & $11$ & $2.3$\\
$3.5$ & $44$   & $11$ & $52$ & $18$ & $3.8$ & $0.44$ & $7.6$ 
& $22$   & $3.0$ & $6.7$ & $1.7$\\
$4.0$ & $47$   & $7.4$ & $31$ & $11$ & $2.8$ & $0.27$ & $8.8$ 
& $13$   & $8.9$ & $4.2$ & $1.3$\\
$6.0$ & $30$   & $1.9$ & $4.8$ & $2.2$ & $0.94$ & $0.04$ & $8.5$ 
& $2.5$   & $22$ & $0.79$ & $0.48$\\
$7.0$ & $22$   & $1.1$ & $2.1$ & $1.1$ & $0.61$ & $0.01$ & $7.7$ 
& $1.2$   & $$22 & $0.38$ & $0.32$\\
$9.0$ & $12$   & $0.47$ & $0.50$ & $0.30$ & $0.31$ & $0.001$ & $6.3$ 
& $0.36$   & $21$ & $0.09$ & $0.17$\\
\hline
Exp. & $227$ & $66$ & $67$ & $109$ & $68$ & $10$ & $15$
 & $119$ & $61$ & $38$ & $45$ \\
 \noalign{\smallskip}\hline
\end{tabular}
\end{center}
\end{table*}
\end{document}